\documentclass[submission, Proceedings]{SciPost}

\usepackage{xspace} 
\usepackage{upgreek}

\def\lhcb   {\mbox{LHCb}\xspace}
\def\atlas  {\mbox{ATLAS}\xspace}
\def\cms    {\mbox{CMS}\xspace}

\def\lhc    {\mbox{LHC}\xspace}

\def\MagUp {\mbox{\em Mag\kern -0.05em Up}\xspace}

 \def\Ppi         {\ensuremath{\uppi}\xspace}

 \def\Pphi        {\ensuremath{\upphi}\xspace}                 
                  
 \def\Pchi        {\ensuremath{\upchi}\xspace}                 
 \def\Ppsi        {\ensuremath{\uppsi}\xspace}

 \def\PDelta      {\ensuremath{\Delta}\xspace}                 
 \def\PXi         {\ensuremath{\Xi}\xspace}                 
 \def\PLambda     {\ensuremath{\Lambda}\xspace}                 
 \def\PSigma      {\ensuremath{\Sigma}\xspace}                 
 \def\POmega      {\ensuremath{\Omega}\xspace}                 
 \def\PUpsilon    {\ensuremath{\Upsilon}\xspace}

 \def\PB      {\ensuremath{\mathrm{B}}\xspace}                 
                  
 \def\PD      {\ensuremath{\mathrm{D}}\xspace}

 \def\PJ      {\ensuremath{\mathrm{J}}\xspace}                 
 \def\PK      {\ensuremath{\mathrm{K}}\xspace}

 \def\Pb      {\ensuremath{\mathrm{b}}\xspace}

 \def\Pi      {\ensuremath{\mathrm{i}}\xspace}

 \def\Pp      {\ensuremath{\mathrm{p}}\xspace}

 \def\Ps      {\ensuremath{\mathrm{s}}\xspace}

 \def\thebaroffset{0.0em}

 \def\Ppi         {\ensuremath{\pi}\xspace}

 \def\Pphi        {\ensuremath{\phi}\xspace}                 
                  
 \def\Pchi        {\ensuremath{\chi}\xspace}                 
 \def\Ppsi        {\ensuremath{\psi}\xspace}                 
                  
 \mathchardef\PDelta="7101
 \mathchardef\PXi="7104
 \mathchardef\PLambda="7103
 \mathchardef\PSigma="7106
 \mathchardef\POmega="710A
 \mathchardef\PUpsilon="7107
                  
 \def\PB      {\ensuremath{B}\xspace}                 
                  
 \def\PD      {\ensuremath{D}\xspace}

 \def\PJ      {\ensuremath{J}\xspace}                 
 \def\PK      {\ensuremath{K}\xspace}

 \def\Pb      {\ensuremath{b}\xspace}

 \def\Pi      {\ensuremath{i}\xspace}

 \def\Pp      {\ensuremath{p}\xspace}

 \def\Ps      {\ensuremath{s}\xspace}

 \def\thebaroffset{0.18em}

\newcommand{\offsetoverline}[2][\thebaroffset]{\kern #1\overline{\kern -#1 #2}}%

\makeatletter
\ifcase \@ptsize \relax
  \newcommand{\miniscule}{\@setfontsize\miniscule{4}{5}}
\or
  \newcommand{\miniscule}{\@setfontsize\miniscule{5}{6}}
\or
  \newcommand{\miniscule}{\@setfontsize\miniscule{5}{6}}
\fi
\makeatother

\DeclareRobustCommand{\optbar}[1]{\shortstack{{\miniscule (\rule[.5ex]{1.25em}{.18mm})}
  \\ [-.7ex] $#1$}}

\def\squark    {{\ensuremath{\Ps}}\xspace}

\def\bquark    {{\ensuremath{\Pb}}\xspace}

\def\pion   {{\ensuremath{\Ppi}}\xspace}

\def\pip    {{\ensuremath{\pion^+}}\xspace}
\def\pim    {{\ensuremath{\pion^-}}\xspace}

\def\kaon    {{\ensuremath{\PK}}\xspace}

\def\KorKbar {\kern \thebaroffset\optbar{\kern -\thebaroffset \PK}{}\xspace}

\def\Kp      {{\ensuremath{\kaon^+}}\xspace}
\def\Km      {{\ensuremath{\kaon^-}}\xspace}

\def\Kstarz  {{\ensuremath{\kaon^{*0}}}\xspace}

\newcommand{\phiz}{\ensuremath{\Pphi}\xspace}

\def\D       {{\ensuremath{\PD}}\xspace}

\def\DorDbar {\kern \thebaroffset\optbar{\kern -\thebaroffset \PD}\xspace}

\def\Dp      {{\ensuremath{\D^+}}\xspace}
\def\Dm      {{\ensuremath{\D^-}}\xspace}

\def\DpDm    {\ensuremath{\Dp {\kern -0.16em \Dm}}\xspace}

\def\B       {{\ensuremath{\PB}}\xspace}

\def\BorBbar {\kern \thebaroffset\optbar{\kern -\thebaroffset \PB}\xspace}

\def\Bd      {{\ensuremath{\B^0}}\xspace}

\def\BdorBdbar {\kern \thebaroffset\optbar{\kern -\thebaroffset \Bd}\xspace}
\def\Bu      {{\ensuremath{\B^+}}\xspace}

\def\Bs      {{\ensuremath{\B^0_\squark}}\xspace}

\def\BsorBsbar {\kern \thebaroffset\optbar{\kern -\thebaroffset \Bs}\xspace}

\def\jpsi     {{\ensuremath{{\PJ\mskip -3mu/\mskip -2mu\Ppsi}}}\xspace}
\def\psitwos  {{\ensuremath{\Ppsi{(2S)}}}\xspace}

\def\Y#1S{\ensuremath{\PUpsilon{(#1S)}}\xspace}

\def\theX     {{\ensuremath{\Pchi_{c1}(3872)}}\xspace}

\def\proton      {{\ensuremath{\Pp}}\xspace}

\def\Lz          {{\ensuremath{\PLambda}}\xspace}

\def\LorLbar     {\kern \thebaroffset\optbar{\kern -\thebaroffset \PLambda}\xspace}
\def\Lambdares   {{\ensuremath{\PLambda}}\xspace}

\def\Xires       {{\ensuremath{\PXi}}\xspace}

\def\Xim         {{\ensuremath{\Xires^-}}\xspace}

\def\Lb           {{\ensuremath{\Lz^0_\bquark}}\xspace}

\def\Xibm         {{\ensuremath{\Xires^-_\bquark}}\xspace}

\def\BF         {{\ensuremath{\mathcal{B}}}\xspace}

\def\to                 {\ensuremath{\rightarrow}\xspace}

\def\AT#1     {\ensuremath{A_{\mathrm{T}}^{#1}}\xspace}           

\def\C#1      {\ensuremath{\mathcal{C}_{#1}}\xspace}                       
\def\Cp#1     {\ensuremath{\mathcal{C}_{#1}^{'}}\xspace}                    
\def\Ceff#1   {\ensuremath{\mathcal{C}_{#1}^{\mathrm{(eff)}}}\xspace}        
\def\Cpeff#1  {\ensuremath{\mathcal{C}_{#1}^{'\mathrm{(eff)}}}\xspace}       
\def\Ope#1    {\ensuremath{\mathcal{O}_{#1}}\xspace}                       
\def\Opep#1   {\ensuremath{\mathcal{O}_{#1}^{'}}\xspace}                    

\newcommand{\aunit}[1]{\ensuremath{\text{\,#1}}}       

\newcommand{\tev}{\aunit{Te\kern -0.1em V}\xspace}
\newcommand{\gev}{\aunit{Ge\kern -0.1em V}\xspace}
\newcommand{\mev}{\aunit{Me\kern -0.1em V}\xspace}
\newcommand{\kev}{\aunit{ke\kern -0.1em V}\xspace}
\newcommand{\ev}{\aunit{e\kern -0.1em V}\xspace}
 
\newcommand{\mevc}{\ensuremath{\aunit{Me\kern -0.1em V\!/}c}\xspace}
\newcommand{\gevc}{\ensuremath{\aunit{Ge\kern -0.1em V\!/}c}\xspace}
\newcommand{\mevcc}{\ensuremath{\aunit{Me\kern -0.1em V\!/}c^2}\xspace}
\newcommand{\gevcc}{\ensuremath{\aunit{Ge\kern -0.1em V\!/}c^2}\xspace}

\def\fb   {\ensuremath{\aunit{fb}}\xspace}
\def\invfb   {\ensuremath{\fb^{-1}}\xspace}

\newcommand{\stat}{\aunit{(stat)}\xspace}
\newcommand{\syst}{\aunit{(syst)}\xspace}

\newcommand{\chisq}{\ensuremath{\chi^2}\xspace}

\def\gsim{{~\raise.15em\hbox{$>$}\kern-.85em
          \lower.35em\hbox{$\sim$}~}\xspace}
\def\lsim{{~\raise.15em\hbox{$<$}\kern-.85em
          \lower.35em\hbox{$\sim$}~}\xspace}

\def\tell1  {TELL1\xspace}
\def\ukl1   {UKL1\xspace}

\newcommand{\etal}{\mbox{\itshape et al.}\xspace}

\usepackage{times} 
\usepackage{url}
\usepackage{amsmath}
\usepackage{float}
\usepackage{mathtools}
\usepackage{caption}
\usepackage{graphbox}
\usepackage{graphicx, subcaption}
\usepackage{textcmds}
\usepackage{ifthen} 
\newboolean{uprightparticles}
\setboolean{uprightparticles}{false} 
\usepackage{xspace} 
\usepackage{upgreek}
\usepackage{fancyhdr}

\hypersetup{
    colorlinks,
    linkcolor={red!50!black},
    citecolor={blue!50!black},
    urlcolor={blue!80!black}
}

\usepackage[bitstream-charter]{mathdesign}
\DeclareSymbolFont{usualmathcal}{OMS}{cmsy}{m}{n}
\DeclareSymbolFontAlphabet{\mathcal}{usualmathcal}

\begin{document}

\begin{center}{\Large \textbf{
Recent Studies of Exotic Hadrons at the LHCb Experiment\\
}}\end{center}

\begin{center}
G. Robertson\textsuperscript{1*}
\end{center}

\begin{center}
{\bf 1} School of Physics and Astronomy, University of Edinburgh, Edinburgh, Scotland
\\
On behalf of the LHCb Collaboration
\\
* gary.robertson@cern.ch
\end{center}

\begin{center}
\today
\end{center}


\definecolor{palegray}{gray}{0.95}
\begin{center}
\colorbox{palegray}{
  \begin{tabular}{rr}
  \begin{minipage}{0.1\textwidth}
    \includegraphics[width=22mm]{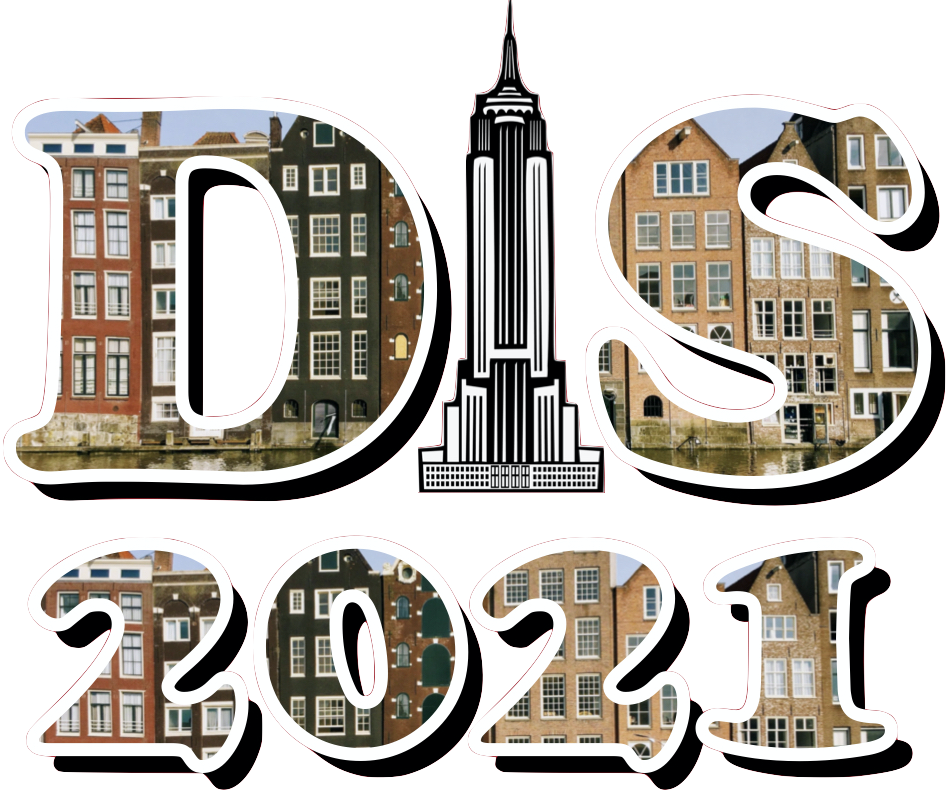}
  \end{minipage}
  &
  \begin{minipage}{0.75\textwidth}
    \begin{center}
    {\it Proceedings for the XXVIII International Workshop\\ on Deep-Inelastic Scattering and
Related Subjects,}\\
    {\it Stony Brook University, New York, USA, 12-16 April 2021} \\
    \doi{10.21468/SciPostPhysProc.?}\\
    \end{center}
  \end{minipage}
\end{tabular}
}
\end{center}

\fancypagestyle{noheader}{
  \fancyhf{}
  \renewcommand{\headrulewidth}{0pt}
}
\ifnum\value{page}=1 \thispagestyle{noheader}%
\section*{Abstract}
{\bf
Recent results on studies of exotic hadrons at the LHCb experiment are summarised. These are the observation of new resonances decaying to $\jpsi\Kp$ and $\jpsi\phiz$, evidence of a $\jpsi\Lambdares$ structure and observation of excited \Xim states in the $\Xibm\to\jpsi\Lambdares\Km$ decay and the study of $\Bs\to\jpsi\pip\pim\Kp\Km$ decays.
}


\section{Introduction}
\label{sec:intro}
Exotic hadrons - defined as hadrons with more than 3 valence quarks - were first proposed by Gell-Mann himself in his paper on the quark model \cite{Gell_Mann}. Their existence was only recently confirmed in 2003 by the Belle collaboration who reported the existence of an exotic hadron named the $\theX$ found in the $\jpsi\pip\pim$ mass spectrum \cite{belle}. Pentaquarks, which are states with 5 valence quarks, were first observed in 2015 by the \lhcb collaboration in the $\jpsi\proton$ mass spectrum, and then most recently observed in 2019 with a larger dataset \cite{pentaquarks2015,pentaquarks2019}.

Since the \lhc started collecting data \atlas, \cms and \lhcb have discovered 59 new hadronic states, the most recent of which was the observation of four new tetraquarks by \lhcb \cite{Paper1} which will be discussed in the following section.


\section{Observation of New Resonances Decaying to $\jpsi\Kp$ and $\jpsi\phiz$.}
\label{sec:tetraquarks}

The $\Bu\to\jpsi\phiz\Kp$ mode has been previously studied by \lhcb using only the run 1 dataset, corresponding to 3\invfb of data \cite{Paper1prev1, Paper1prev2}. In these studies four new $X\to\jpsi\phiz$ structures were observed at greater than 5$\sigma$ significance, labelled as the X(4140), X(4274), X(4500) and X(4700) states. They are highlighted in the plot shown in figure \ref{fig:paper1fig12}.

\begin{figure}[!htb]
    \centering
    \includegraphics[width=0.45\linewidth]{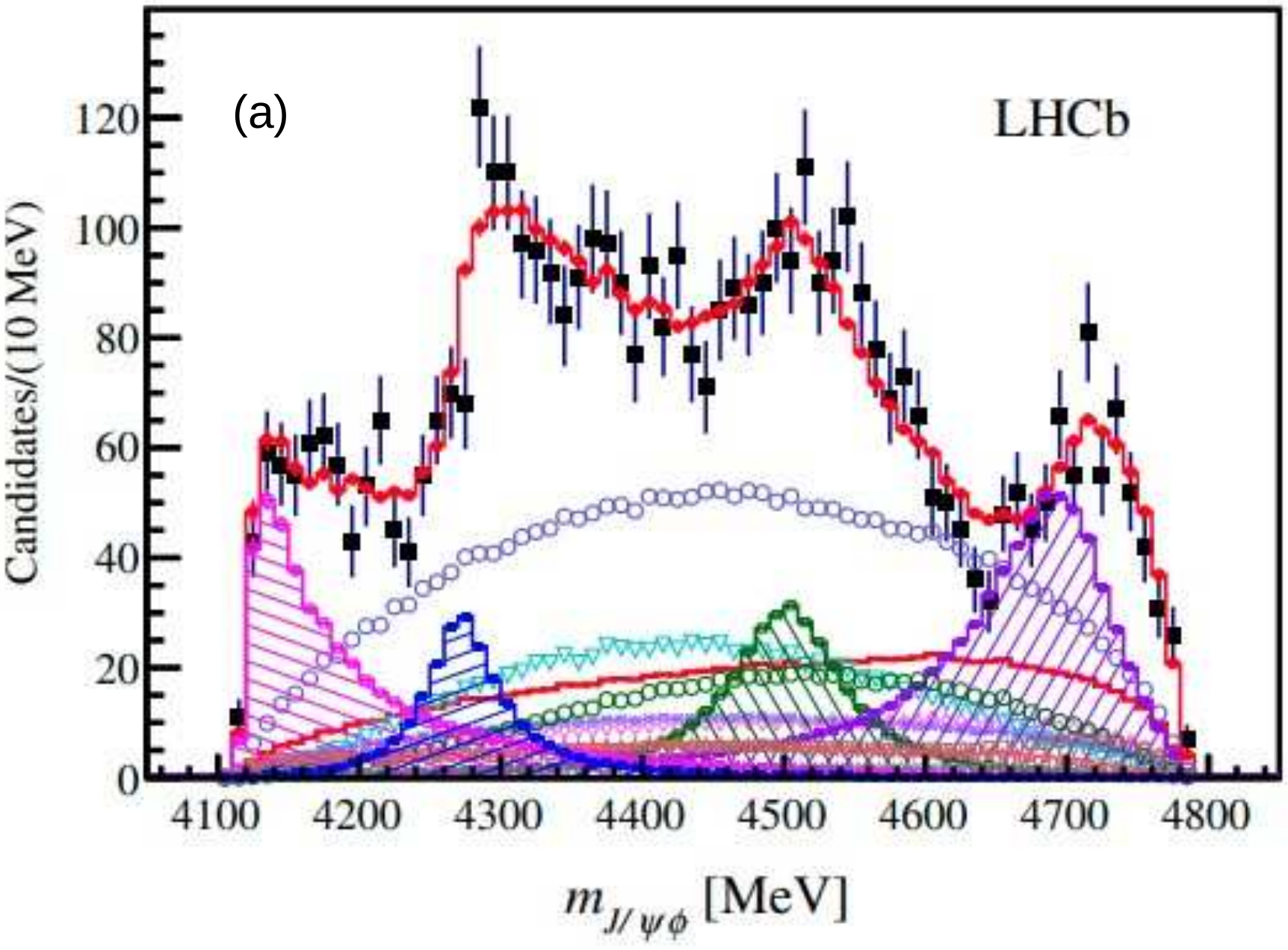}
    \includegraphics[width=0.48\linewidth]{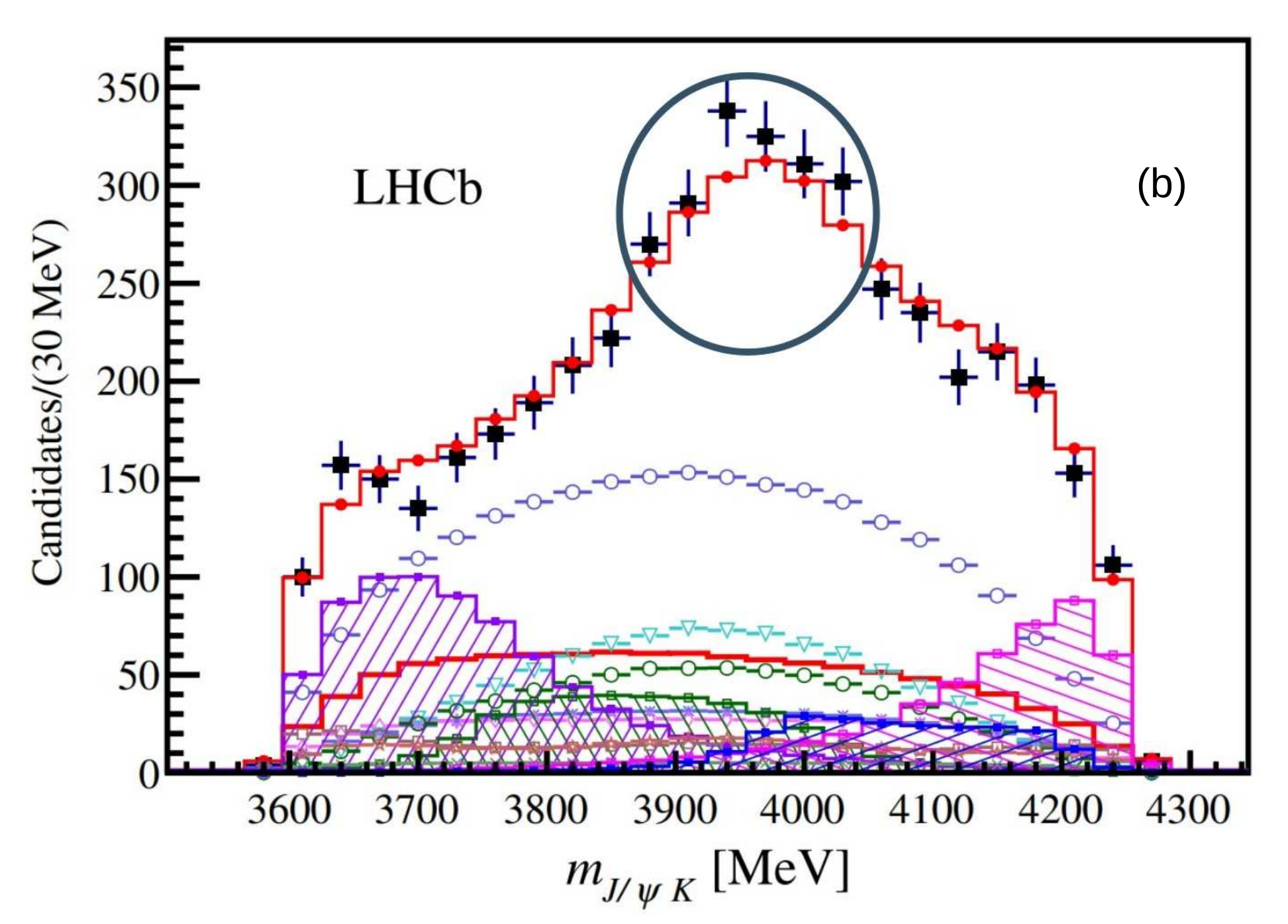}
    \caption{\textit{(a) The $\jpsi\phiz$ mass spectrum is shown. The multiple shaded histograms show the contributions from each of the X states, X(4140) , X(4274), X(4500) and X(4700)} respectively. \textit{(b) The $\jpsi\kaon$ mass spectrum is shown. The circle highlights the area where there is a clear discrepancy between the data and the fit. This was measured to be 3$\sigma$.} \textit{In both plots the data are represented by the black boxes, with the fit is represented by the red line.}}
    \label{fig:paper1fig12}
\end{figure}

In the $\jpsi\Kp$ mass spectrum, a peak around 4\gev was also observed, as shown in figure~\ref{fig:paper1fig12}. Its significance was measured as 3$\sigma$ so was not large enough to claim an observation but was suggested to be a hint of a possible $Z_{cs}^+$ state.

In this new analysis, full advantage is taken of the Run 1 and 2 dataset - corresponding to 9\invfb \cite{Paper1}. There were also some small changes to the preselection; such as a looser vertex-fit $\chisq$ value for the kaons. The machine learning algorithms in use were improved, leading to better background subtraction. The combination of these factors led to 6 times more signal yield (about 24,000 candidates) and 6 times less combinatorial background.

In Dalitz plots the X structures that were previously reported were seen very clearly in the $\jpsi\phiz$ spectrum, and a structure was observed in the $\jpsi\Kp$ spectrum, as shown in figure \ref{fig:paper1fig3}. A full amplitude analysis was then carried out.

\begin{figure}[!htb]
    \centering
    \includegraphics[width=0.48\linewidth]{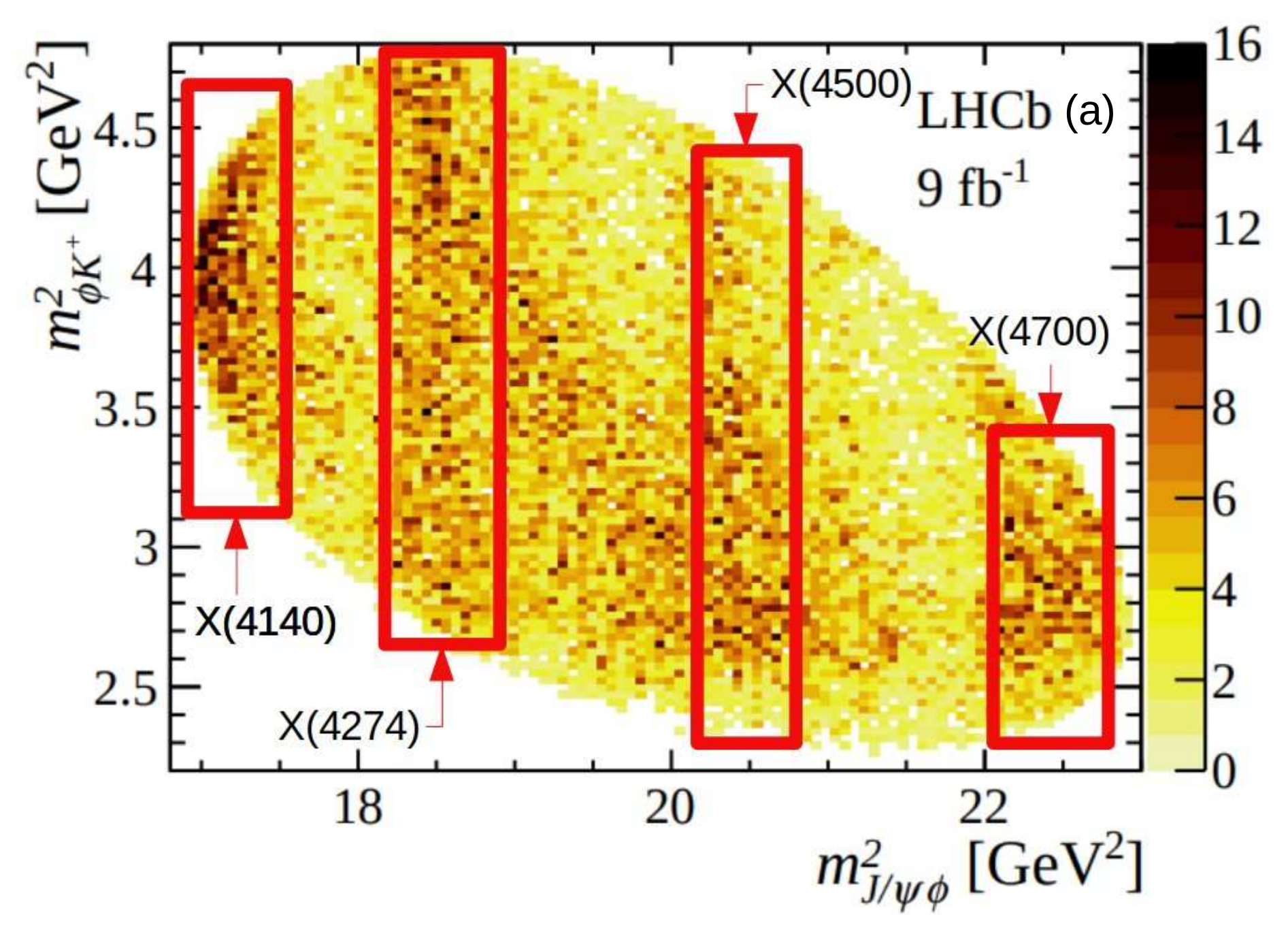}
    \includegraphics[width=0.49\linewidth]{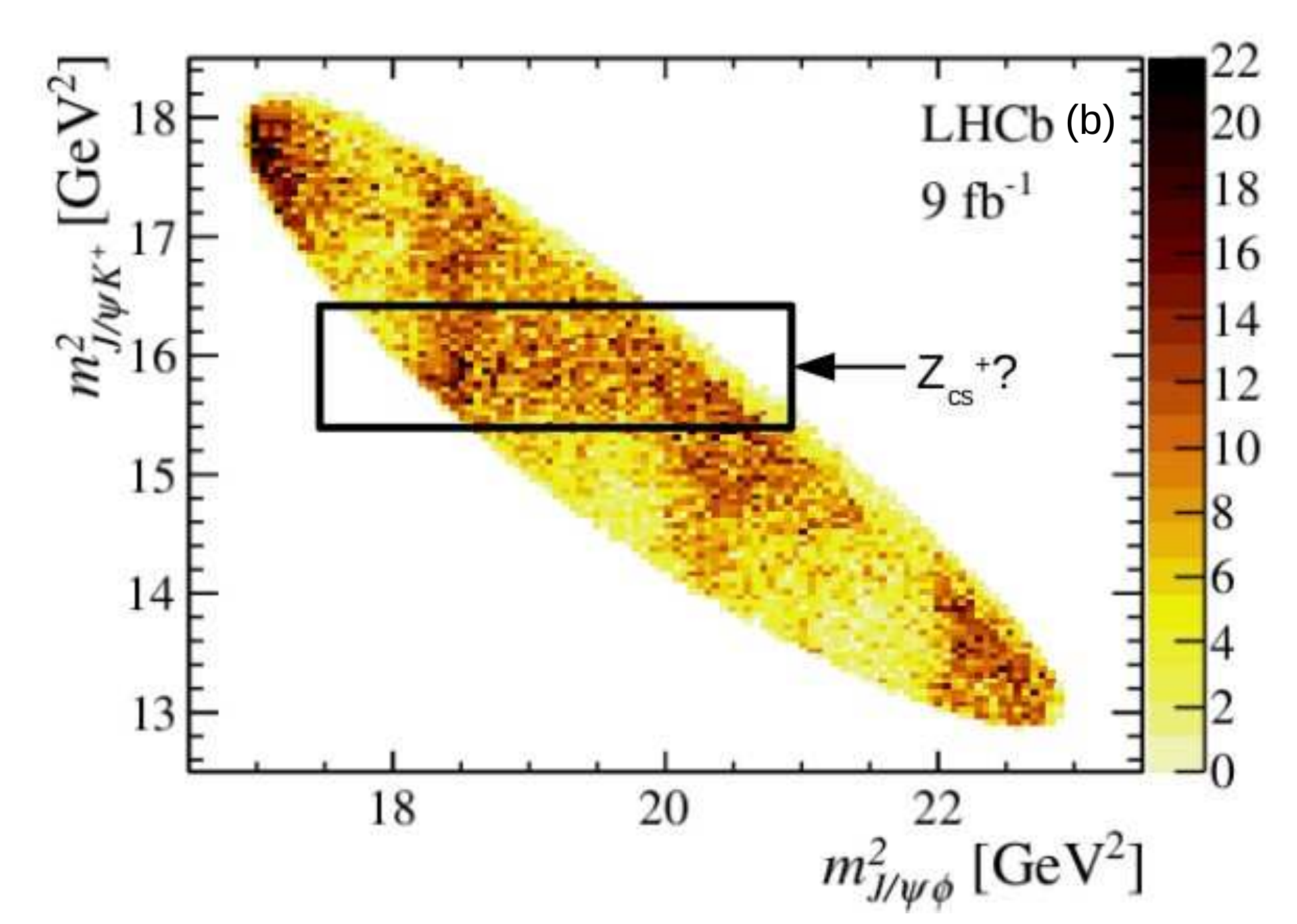}
    \caption{\textit{(a) The Dalitz plot of the $\jpsi\phiz$ squared mass against the $\phiz\Kp$ squared mass. The bands corresponding to the X structures are highlighted. (b) The Dalitz plot of the $\jpsi\phiz$ squared mass against the $\jpsi\Kp$ squared mass. A band which could possibly come from a $Z_{cs}^+$ structure is highlighted.}}
    \label{fig:paper1fig3}
\end{figure}

In the fit the four X states previously observed as well as two new X states and two $Z_{cs}$ states were included, shown in figure \ref{fig:paper1fit}. The four established X states were found to be consistent with the previous study, and two new X states labelled as X(4630) and X(4685) were found. In particular X(4685) was found with high significance. Two new $Z_{cs}^+$ states were found in the $\jpsi\Kp$ mass spectrum, labelled as $Z_{cs}(4000)^+$ and $Z_{cs}(4220)^+$.

The spin parity of these states was measured, and for the X(4685) it was found to be $1^+$ with a high significance, and for X(4630) either $1^-$ or $2^-$ at greater than $5\sigma$ significance. For the $Z_{cs}(4000)^+$ it was measured to be $1^+$ with high significance, and for $Z_{cs}(4220)^+$ either $1^+$ or $1^-$ at higher than $5\sigma$ significance.

\begin{figure}[!htb]
    \centering
    \includegraphics[width=\linewidth]{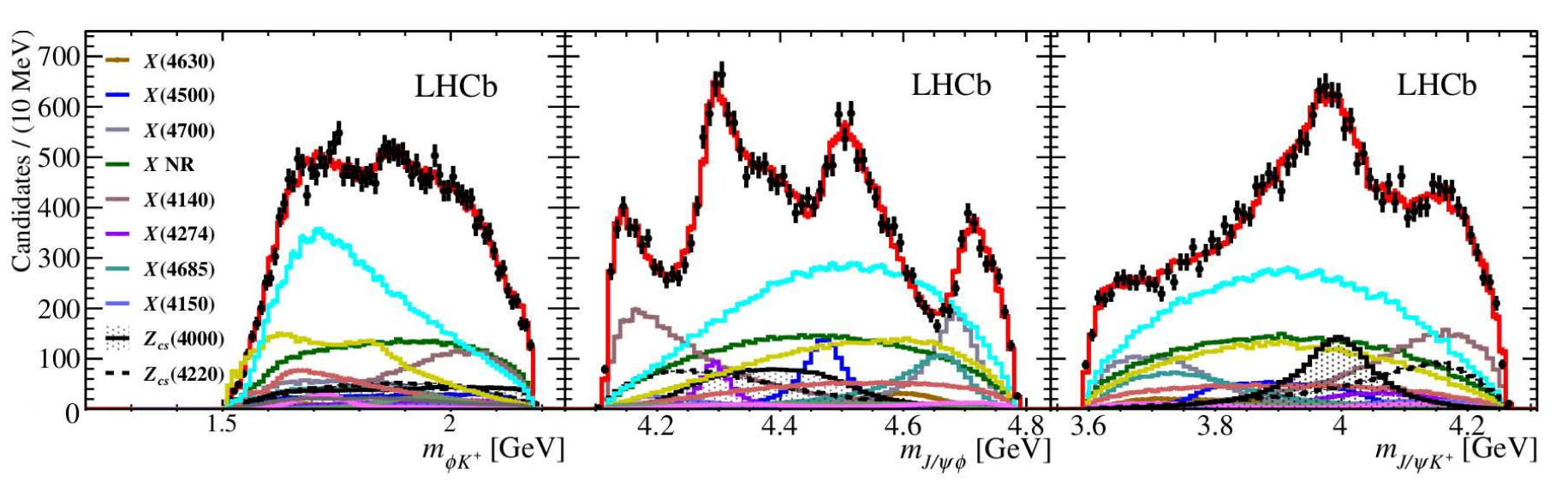}
    \caption{\textit{The fit is shown for three different mass pairs; $\phiz\Kp$, $\jpsi\phi$, and $\jpsi\Kp$ from left to right respectively. The contributions from each of the resonances is shown, with the total fit shown by the solid red line}.}
    \label{fig:paper1fit}
\end{figure}
\section{Evidence of a $\jpsi\Lambdares$ Structure and Observation of Excited \Xim States in the $\Xibm\to\jpsi\Lambdares\Km$ Decay.}

In 2019, the $\jpsi\proton$ mass spectrum was studied once more, paying particular attention to the area around the pentaquark masses. This is when $P_c(4312)^+$ was observed, and $P_c(4450)^+$ was resolved into two new states, labelled as $P_c(4440)^+$ and $P_c(4457)^+$. By swapping one of the charm quarks in the proton for a strange quark, the strange counterparts $P_{cs}$ can be searched for, as seen in reference \cite{Paper2}. These strange charmed pentaquarks were predicted to be found in $\Xibm$ decays in 2010 \cite{Pcs2010}.
%

In the Dalitz plot shown in figure \ref{fig:P2Dalitz}, two resonances in the $\Lambda\Km$ mass spectrum can be seen which correspond to the $\Xi(1690)^-$ and $\Xi(1820)^-$ states respectively. An amplitude analysis was then carried out to measure the properties of the $\Xi$ states and to look for any contributions from $P_{cs}^0$ pentaquark states.

\begin{figure}[!htb]
    \centering
    \includegraphics[width=0.5\linewidth]{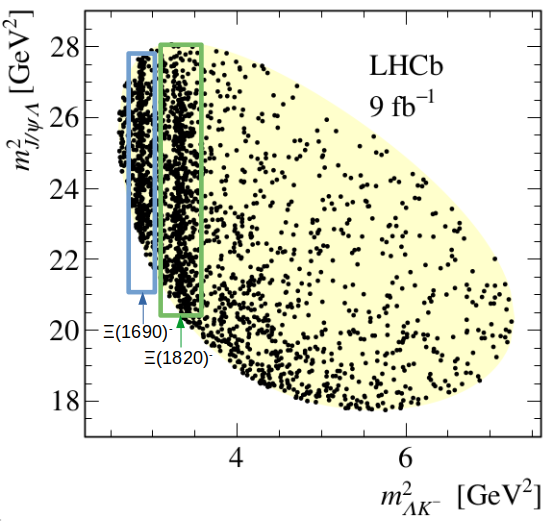}
    \caption{\textit{The squared mass of the $\Lambda\Km$ is plotted against the squared mass of the $\jpsi\Lambda$. Two vertical bands are highlighted which correspond to resonances from two different $\Xi$ states}.}
    \label{fig:P2Dalitz}
\end{figure}

The amplitude analysis was carried out allowing for two new $\Xi$ states (in this decay mode) as well as one new $P_{cs}^0$ state. The results are shown in figure \ref{fig:p3fig3}. There is a significant contribution from the two $\Xim$ states previously mentioned, and a sizeable contribution from the $P_{cs}$ state, measured as $3.1\sigma$ significance.

\begin{figure}[!htb]
    \centering
    \includegraphics[width=0.47\linewidth]{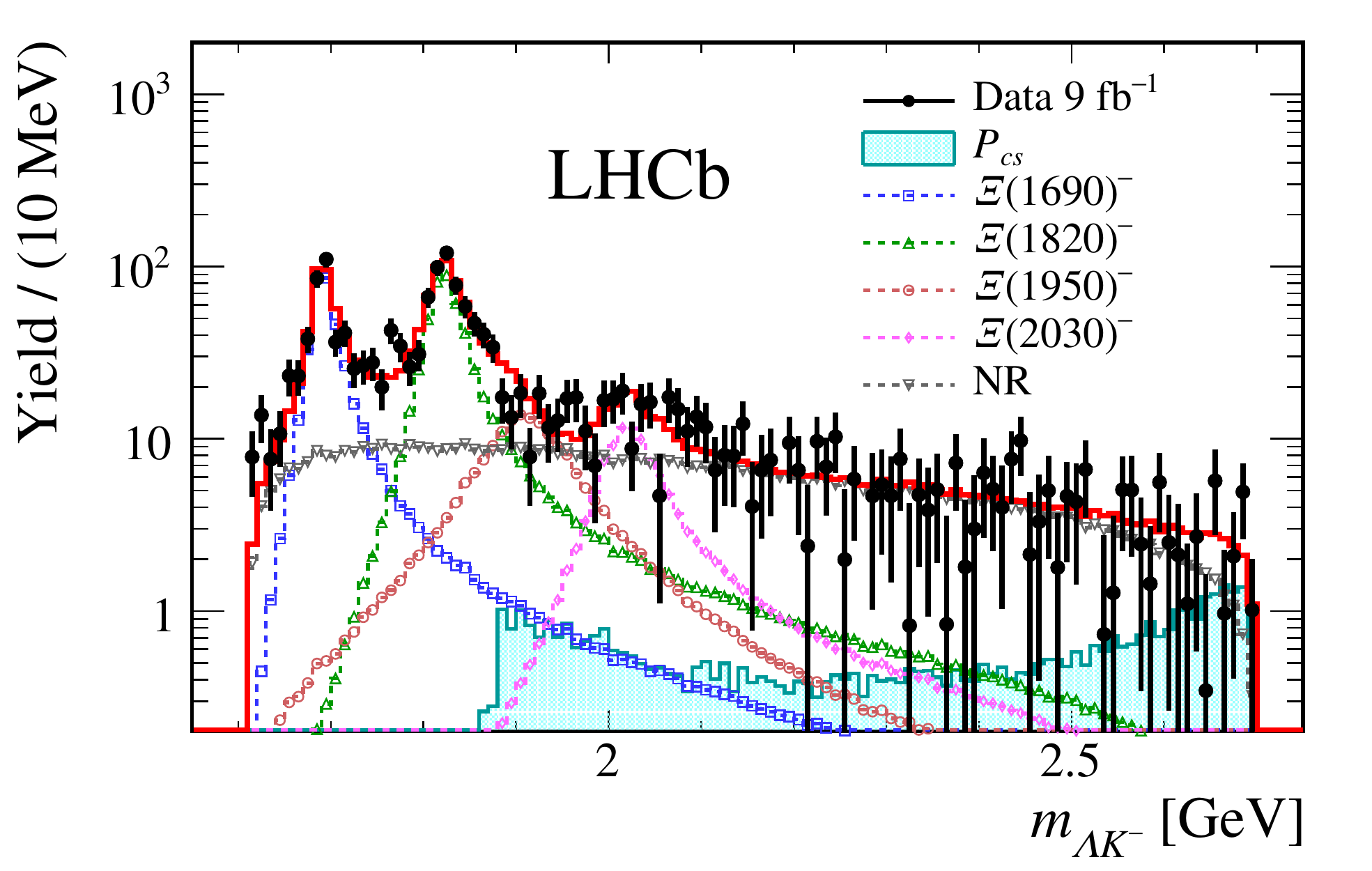}
    \includegraphics[width=0.47\linewidth]{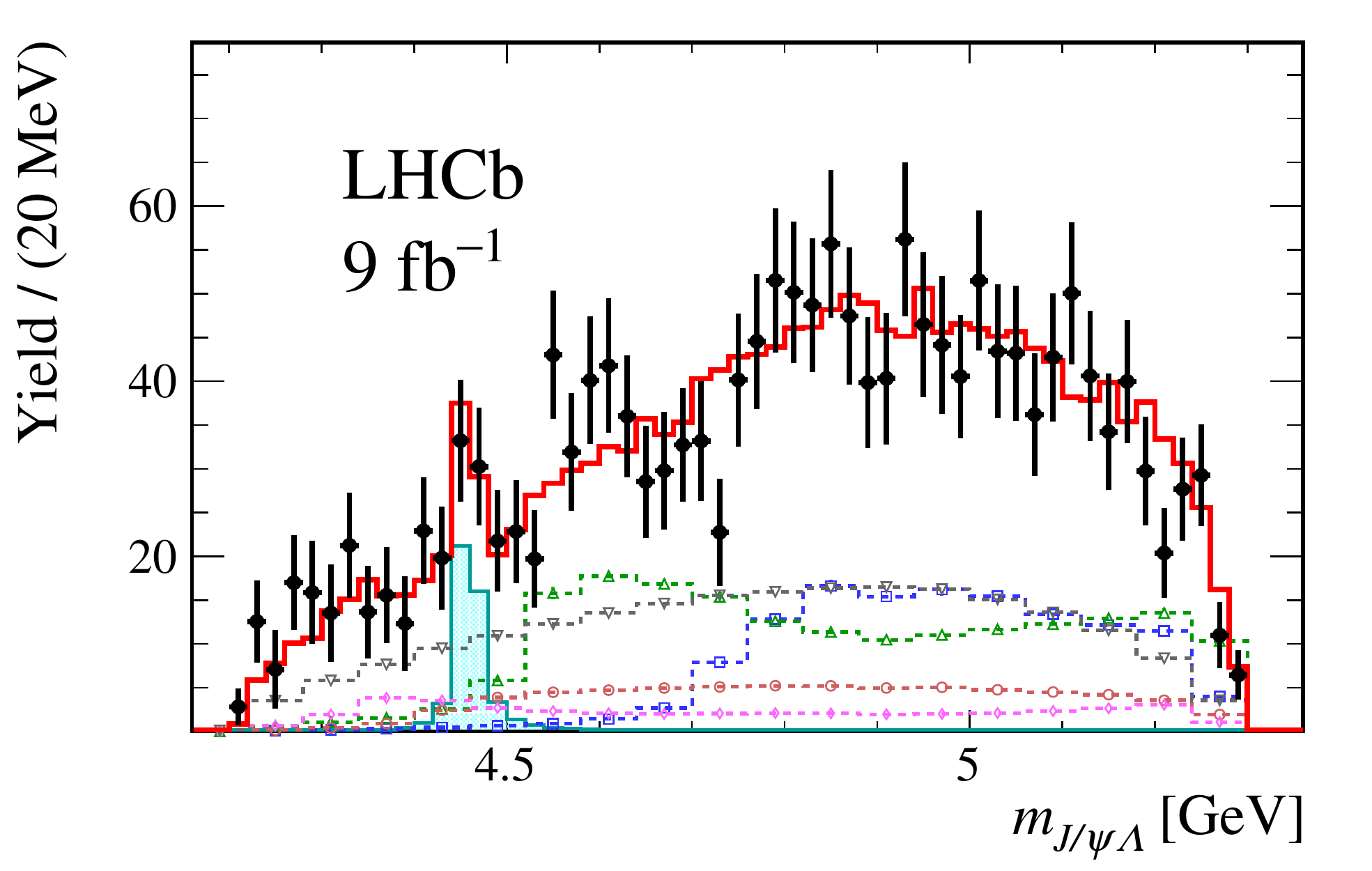}
    \caption{\textit{The left plot shows the fit to the $\Lambda\Km$ mass spectrum and the right to the $\jpsi\Lambda$ spectrum respectively. Contributions from various $\Xim$ states are shown with coloured lines, as well as the contribution from the $P_{cs}$ state shown by the solid blue area. The total fit is represented by the solid red line}.}
    \label{fig:p3fig3}
\end{figure}

The hypothesis that the peak from the $P_{cs}$ state actually consists of two peaks was also tested, but there was not enough data to confirm or refute this.
\section{Study of $\Bs\to\jpsi\pip\pim\Kp\Km$ Decays.}

This decay mode was chosen to be studied as it provides a good channel to look for possible $\jpsi\phiz$ resonances in the $\Bs\to\jpsi\pip\pim\phiz$ decay (about 26,000 candidates). It is also
possible to make measurements of the branching fraction for both the $\Bs\to\theX\phiz$ and $\Bs\to\theX\Km\Kp$ decays \cite{Paper3}.

The three ratios $R_{\psitwos\phiz}^{\theX\phiz}$, $R_{\psitwos\phiz}^{\jpsi\Kstarz\bar{\Kstarz}}$ and $R_{\Kp\Km}$ were also measured, where $R_{\psitwos\phiz}^{\theX\phiz}$ is defined as:
\begin{equation}
\begin{aligned}
    R_{\psitwos\phiz}^{\theX\phiz} \equiv \frac{\BF(\Bs\to\theX\phiz)\times\BF(\theX\to\jpsi\pip\pim)}{\BF(\Bs\to\psitwos\phiz)\times\BF(\psitwos\to\jpsi\pip\pim)\times\BF(\psitwos\to\Kp\Km)},
\end{aligned}
\end{equation}
and the other definitions are analogous.

A structure was also observed in the $\jpsi\phiz$ mass spectrum, shown in figure \ref{fig:P3fig10}. This new state, labelled X(4740), was found with a significance of $5.3\sigma$. This could be the X(4700) state, as mentioned in section \ref{sec:tetraquarks}, but more data are required in order to accept or refute this.

\begin{figure}[!htb]
    \centering
    \includegraphics[width=0.6\linewidth]{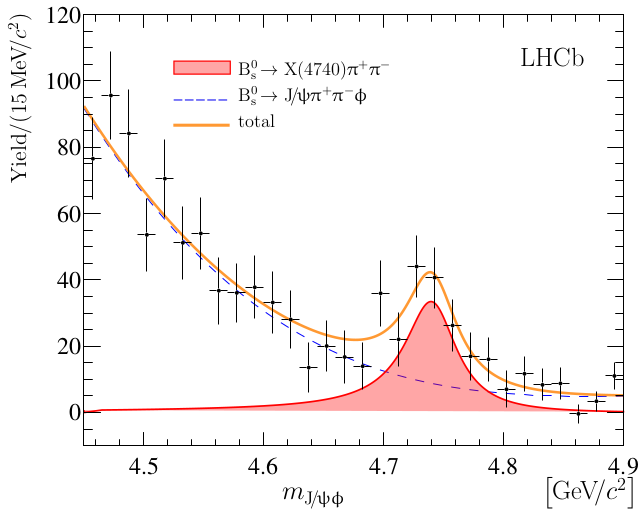}
    \caption{{\it The $\jpsi\phiz$ mass spectrum is shown, where the red shaded area shows the contribution of the X(4740) state, the blue dashed line represents the background, and the orange solid line shows the total fit}.}
    \label{fig:P3fig10}
\end{figure}

The $\Bs$ mass was also measured using the $\psitwos\Kp\Km$ mass distribution enriched in $\Bs\to\psitwos\phiz$ decays. This is the most precise single measurement of the $\Bs$ mass, and was measured as $5366.98 \pm 0.07\stat \pm 0.13\syst\mev$. The ratios are summarised as:
\begin{equation*}
\begin{aligned}
    R_{\psitwos\phiz}^{\theX\phiz} &= (2.42\pm0.23\pm0.07)\times10^{-2},\\ R_{\psitwos\phiz}^{\jpsi\Kstarz\bar{\Kstarz}} &= 1.22\pm0.03\pm0.04, \\ 
    R_{\Kp\Km} &= 1.57\pm0.32\pm0.12,
\end{aligned}
\end{equation*}
where the first uncertainty is statistical and the second is systematic.
\section{Conclusion}
Many new exotic states have been discussed here, in particular; three new X structures and two new $Z_{cs}$ states, evidence of a charmed pentaquark with strangeness as well as measurements of important branching fractions. However, there are still many results to come from the current \lhcb dataset in the next few years, as well as many more results to come from future runs (for example Run 3 and Run 4 will collect 50\invfb in total). Current and future upgrades will also improve the detector such that analysts are able to fully exploit it's capabilities.

\nolinenumbers

\end{document}